\documentclass[twocolumn,aps,epsf,showpacs]{revtex4}

\usepackage{graphicx}
\usepackage[french]{babel} 
\usepackage{amsmath} 
\usepackage{times} \usepackage{color} \usepackage[ps2pdf,colorlinks=true,linkcolor=webred]{hyperref} 
\definecolor{webred}{rgb}{0.5,0,0}

\begin{document}

\title{Effect of an electric field on an intermittent granular flow}

\author{E. Mersch$^1$, G. Lumay$^{1,2}$, F. Boschini$^{1,3}$ and N. Vandewalle$^1$}

\affiliation{ 
$^{1}$GRASP, Institut de Physique B5, Universit\'e de Li\`ege, B-4000 Li\`ege, Belgium.\\ 
$^{2}$F.R.S.-F.N.R.S., B-1000 Bruxelles, Belgium\\
$^{3}$LCIS/CMI, Institut de Chimie B6, Universit\'e de Li\`ege, B-4000 Li\`ege, Belgium
}

\begin{abstract} 

Granular gravity driven flows of glass beads have been observed  in a silo with a flat bottom. A DC high electric field has been applied perpendicularly to the silo to tune the cohesion. The outlet mass flow has been measured. An image subtraction technique has been applied to visualize the flow geometry and a spatiotemporal analysis of the flow dynamics has been performed. The outlet mass flow is independent of voltage, but a transition from funnel flow to rathole flow is observed. This transition is of probabilistic nature and an intermediate situation exists between the funnel and the rathole situations. At a given voltage, two kinds of flow dynamics can occur : a continuous flow or an intermittent flow. The electric field increases the probability to observe an intermittent flow.


\end{abstract}

\maketitle

\section{Introduction}

Intermittent flows of granular materials address fundamental questions and their study is important for industrial applications \cite{jenike}, especially in the case of silos. Intermittences are observed in a variety of configurations, including gravity driven flows in vertical pipes \cite{spatio,bertho,chen,wang}, funnel and mass flow in silos \cite{silo_music}, in inclined two-dimensional funnels \cite{inclined,inclined2}, or in hourglasses \cite{sablier1,sablier2,sablier3}, as well as in rotating drums \cite{forsyth}. It also happens when slowly driving a granular material confined in a vertical column \cite{clement,olvarez}. The origins of these intermittences are diverse. A proposed mechanism invokes a vertical gradient of air pressure that periodically blocks the flowing granular material. In a vertical column \cite{spatio,bertho}, this gradient of pressure is due to viscous drag of air by the flowing grains. In hourglasses \cite{sablier1,sablier2}, the gradient of pressure is due to the variation of available air volume in the two chambers when the grains pass from one chamber to the other one. In inclined two-dimensional funnels \cite{inclined}, the narrowing of the flowing zone near the outlet was invoked to explain a periodic densification and blocking of the granular media. A front of  decompaction propagates upward and is followed by the upward propagation of a blockage front. In the case of granular flow through a vertical pipe, intermittences can be induced by the presence of a non-homogeneous electric field near the outlet \cite{chen}. Cohesive arches appear near the outlet and are periodically broken. In slowly driven granular columns \cite{clement,olvarez}, the aging of frictional forces explains a stick/slip motion of the packing. Stick/slip motion has also been invoked to model some intermittences in silos \cite{silo_music}. In many situations \cite{clement,spatio, bertho,olvarez,inclined,forsyth}, the relative humidity of air is an important parameter, but its role is poorly understood. The role of the cohesion on the intermittences is still unknown.

In order to study the effects of cohesion from an experimental point of view, it can be helpful to tune the cohesion of a granular material. This can either be achieved by changing the liquid content of a wet granular material \cite{huang}, by tuning the relative humidity of air \cite{forsyth}, or by applying a magnetic field on a ferromagnetic granular material \cite{geoffroy,geoffroy2,pilling,forsyth,lemaire}.

In the present work, we choose to study the flow of a glass powder through a silo and we tuned the cohesion by applying an electric field. We focused our attention on the evolution of the mean flow and on the shape of the flowing zone with time and voltage. We performed a spatiotemporal analysis of the vertical flow to characterize the intermittences.

In section \ref{setup}, the experimental setup is described. In section \ref{results}, we present experimental results. Effects of an electric field on a dielectric granular material are discussed in \ref{effects}. We discuss the interactions induced between grains in subsection \ref{effect}. The flow geometry and the mass flow behavior are compared to previous literature in subsection \ref{mass_effect}. The intermittences are discussed in subsection \ref{mechanism}.

\section{Experimental Setup}\label{setup}

\begin{figure}[!ht]
\includegraphics[scale=0.3]{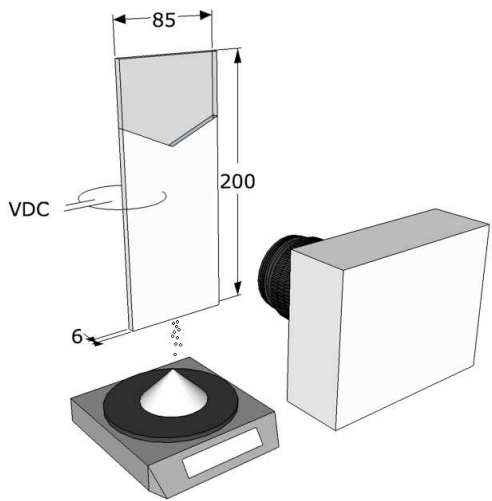}
\caption{A rectangular silo of dimensions 85 mm x 200 mm x 6 mm contains a glass powder. Two vertical plates are plugged to a high voltage power supply to produce a horizontal homogeneous electric field. Air hygrometry is fixed to 43 \%RH. The mass flow is measured with an electronic scale and the flow is visualized with a digital photo camera.}
\label{dispo}
\end{figure}

Figure \ref{dispo} presents a sketch of our experimental setup. We use approximately spherical glass beads of diameter $d$ = 74 $\mu$m with a standard deviation of 7.6 $\mu$m. We use a rectangular silo of dimensions 85 mm x 200 mm x 6 mm with walls made of glass plates. The inner surface of the front and back plates is coated with fluorine-doped tin oxide (FTO), a transparent conductive material. These plate electrodes are connected to a high voltage DC power supply. Voltage is varied from 0 to 1900 V. The outlet of the silo is a rectangular centered slit of 6 mm x 3 mm (i.e. 79 $d$ x 39 $d$) aligned in the direction of the electric field. 
Pictures are either taken with a  fast camera for high temporal resolution or with a digital photo camera for high spatial resolution.

In order to ensure the reproducibility of our measurements, the following procedure has been established. The powder and the silo are placed in a glove box in which the relative humidity is fixed to a value of $43 \%$RH \cite{RH} by the presence of a saturated solution of $K_{2}CO_{3}$. At this hygrometry, one expects to obtain poor cohesion due to the humidity or due to the charges acquired by the triboelectric effect. It is also crucial to control the humidity to avoid fluctuations of the electrical conductivity of the glass beads. The powder is separated in eight samples contained into eight glass vessels. Each sample is used at most once in 24 hours to let the powder relax to its equilibrium under air humidity. The silo is first filled with one sample. We fill it through a glass hopper fixed at the top of the silo. At this moment, the electrodes are both connected to the ground.  After the filling, we wait 20 minutes before applying a voltage. A delay of 20 more minutes is awaited before starting the flow.

The outpoured mass $m(t)$ is measured via an electronic scale controlled by a computer. The sampling rate of the mass measurement is 2 Hz. The response time of the scale is too long to reveal some flow intermittences. The mean mass flow $Q$ is obtained by fitting the mass curve with a linear function. We carried out one measurement for 11 different voltage values between 0 and 1700 V.

Two types of videos are taken. For imaging the geometry of the flowing zone, we use a digital photo camera with a resolution of 2592 x 3888 pixels (167 aligned pixels correspond to 10 mm). The frame rate is 2 images per second. 
To visualize the flowing zone, we subtract each pair of successive images. A grain that appears or disappears from one image to the following one gives a white point in the final image. Blocked and flowing zones respectively appear as dark and white zones in the final pictures. To visualize the dynamics of the flow, we use a fast camera with a resolution of 1024 x 1280 pixels (76 aligned pixels correspond to 10 mm) at a frame rate of 500 images per second. 
We also performed a successive images subtraction for these videos. They were converted to vertical spatiotemporal diagrams to visualize the vertical evolution of the flowing zone with a high temporal resolution.

\section{Experimental Results}\label{results}

\subsection{Mass flow measurements}\label{mass}

\begin{figure}[!ht]
\includegraphics[scale=0.6]{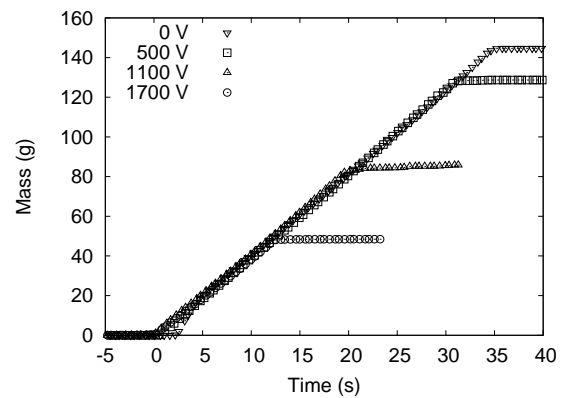} 
\caption{Collected mass during the discharge of the silo for several applied voltages. The slope (i.e. the mass flow $Q$) is constant over time and over voltage. For the highest voltages, the final plateau appears earlier. This reveals that some part of the heap remains blocked in the silo.}
\label{scale}
\end{figure}

Figure \ref{scale} presents four of the eleven mass curves $m(t)$ collected by the electronic scale. A linear evolution of $m(t)$ is reached after less than one second and a horizontal plateau is reached at the end of the discharge. This constant mass flow is expected as a consequence of the saturation of the vertical strain in confined granular materials. Electric field has no effect on $Q$. Its mean value is 4.1 $\pm$ 0.15 g/s.
For the highest voltages, the final plateau appears at a lower level. Only a part of the heap is discharged. We will see in section \ref{geometry} that this is due to a transition of the flow geometry. The represented plots of figure \ref{scale} are for specific drainage runs only. For runs with the same voltage, the level of this plateau would differ from one run to another. Above 1700 V, the heap remains blocked. An arch is observed near the outlet. The flow occurs below this arch, but does not propagate above it.

\subsection{Flow geometry}\label{geometry}

\begin{figure}[!ht]
\includegraphics[scale=0.38]{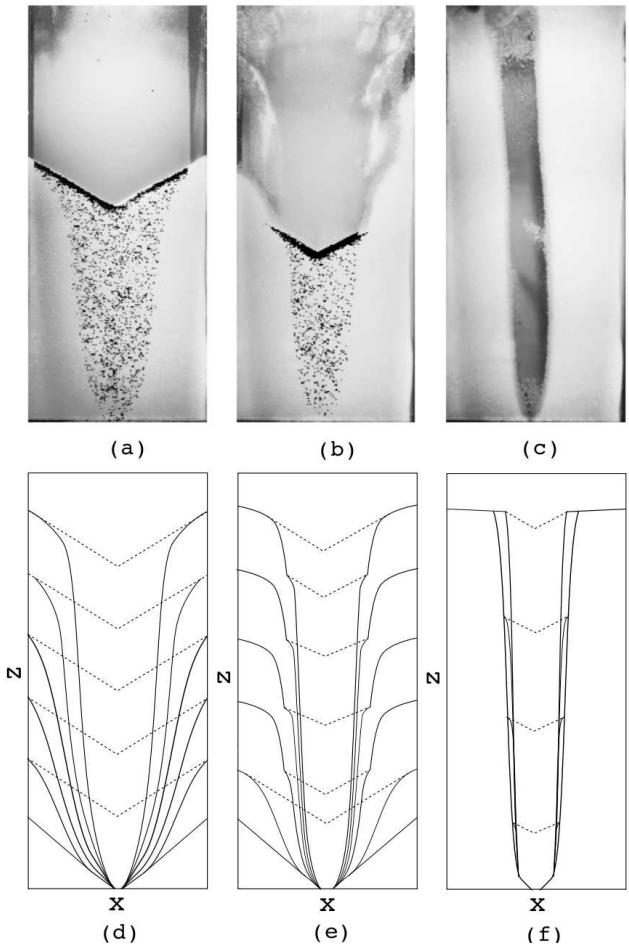} 
\caption{Three kinds of flow geometries are encountered during the silo discharge : (a) funnel flow (0V), (b) avalanche (500 V), (c) rathole (1500 V). The result of the image subtraction has been superposed to the pictures (a) and (b) : black dots correspond to grains that appeared or disappeared from two successive images. Typical temporal evolutions of each regime are sketched on figures (d), (e), (f). The solid lines represent the separation between flowing and non flowing part of the heap. The dashed lines represent the top of the heap.}
\label{shapes}
\end{figure}

We encountered three different kinds of flow geometries: the  "funnel flow" regime, the "avalanche" regime, and the "rathole" regime. These different flow geometries are illustrated in figure \ref{shapes}.

In all situations, the flow is limited to the central part of the heap. In the "funnel flow" (see figure \ref{shapes}.(a),(d)), the top layer of the heap flows over all its width, grains are falling in the direction of the center, and only very small avalanches occur. The evolution with time of the shape of the flowing zone has been observed (a sketch is represented on figure \ref{shapes}.(d)). For funnel flows, its behavior is qualitatively similar to the measures of Sielamowicz and coworkers \cite{piv}. Nedderman \cite{nedderman} applied the kinematic model of Liwiniszyn \cite{litwiniszyn} to describe this kind of evolution, and found similar shape and evolution of the flowing bed except a vanishing angle between his curves and the horizontal direction near the outlet.

In the "avalanche" situation (see figure \ref{shapes}.(b),(e)), stagnant walls can stabilize at the border of the cell, the flow being restricted to the centre of the cell over all its height. After some time, the highest part of these lateral walls tumbles, creating an avalanche. After that, new walls appear at a lower level. Avalanches and intermittences are different phenomena. Intermittences can happen without avalanches and {\it vice versa}. Avalanches only concern the top of the lateral walls, while intermittences concern the central part of the heap, as we will see in section \ref{spatio}.

In the "rathole" situation (see figure \ref{shapes}.(c),(e)), the flow is always restricted to the centre of the heap; stagnant walls remain blocked even after the discharge.

If one still increases the voltage over 1700 V, the packing remains blocked. Only a little part of the packing near the outlet can eventually fall, but an arch prevents the packing to flow thereafter. This situation is called "blocked".

\begin{figure}[!ht]
\begin{center} 
\includegraphics[scale=0.6]{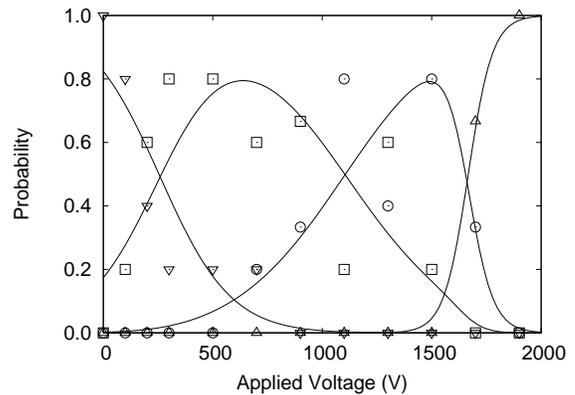} 
\caption{Transitions from funnel flow to blockage. Measured probabilities to observe a funnel flow (reversed triangle), avalanche flow (square), rathole (circle), or blocked flow (triangle) as a function of the applied voltage $V$. The solid lines were obtained by multinomial logistic regression.}
\label{PDF}
\end{center}
\end{figure}

The occurrence of one flow situation is not simply determined by its voltage but rather it happens with some probability. We measured the probability for each of these four situations to happen at different voltages. We have data for twelve different voltage values, and for each voltage, we have at least 5 samples. The result is shown on figure \ref{PDF}. Funnel flows happen more frequently at low voltage, and the probability to observe an avalanche or a rathole increases with the applied voltage. By varying the voltage, the transition from one regime to another one is continuous.

\subsection{Flow dynamics}\label{spatio}
 
\begin{figure}[!ht]
\begin{center} 
\includegraphics[scale=0.6]{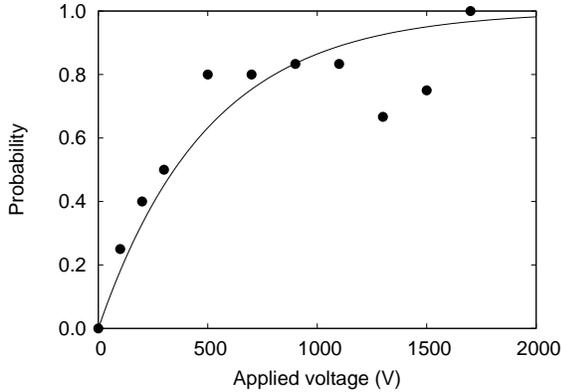} 
\caption{Measured probability to observe an intermittence as a function of the voltage $V$. The saturation curve was obtained by fitting an exponential curve on the data.}
\label{PDF_inter}
\end{center} \end{figure}

\begin{figure}[!ht]
\begin{center} 
\includegraphics[scale=0.32]{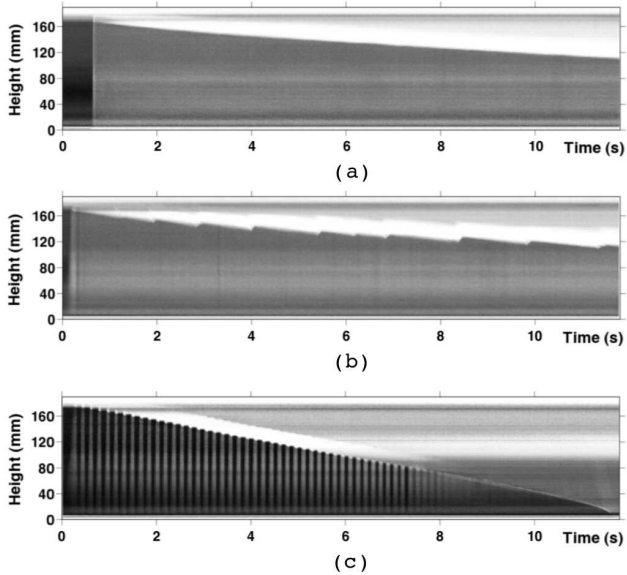} 
\caption{Two types of flow dynamics can be revealed by spatiotemporal analysis of the flow : a continuous flow and an intermittent flow. (a) Continuous flow for a funnel geometry (0 V), (b) Continuous flow with avalanches (1300 V), (c) Intermittent flow in a rathole (900 V). White zones correspond to flowing zones and dark ones to blocked ones.} 
\label{all}
\end{center} \end{figure}

The videos taken with a fast camera have shown that two kinds of flow dynamics can happen at a given voltage: a continuous flow or an intermittent flow.  For 11 different voltage values between 0 and 1700 V, we took at least four videos. The intermittence does not appear systematically, but the chance to observe it increases with the applied voltage (see fig. \ref{PDF_inter}). To characterize this phenomenon, we performed a spatiotemporal analysis of the flow. Successive images were first subtracted to visualize the moving part of the heap (white points correspond to moving grains in the modified images). The gray level of each image is then averaged in the horizontal direction to give an image of 1 pixel width. All the video is then converted to a single spatiotemporal diagram by collating the successive images. Spatiotemporal diagrams can also help to determine the type of flow geometry since they exhibit the evolution of the top of the heap. For funnel flows, its level decreases continuously ((see fig. \ref{all}.(a)). For avalanche flows, the level does not decrease continuously, but the mean slope is the same as in the funnel flow case ((see fig. \ref{all}.(b)). For rathole flow, the level decreases continuously with a greater slope ((see fig. \ref{all}.(c)). The occurrence of avalanche does not imply the occurrence of intermittences (see figure \ref{all}.(b)). When the intermittence is observed, it has a well characterized period $T$.

\begin{figure}[!ht]
\begin{center} 
\includegraphics[scale=0.6]{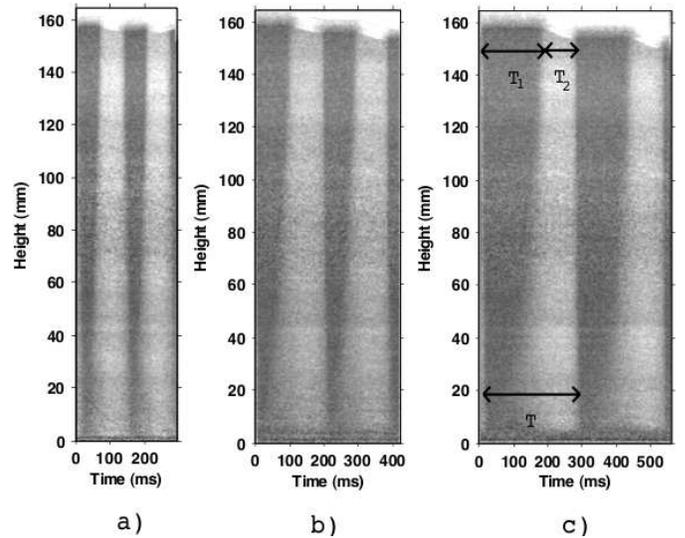} 
\caption{
Zoom on two periods of the intermittence for respectively (a) $V$=300V, (b) $V$=700V, (c) $V$=1500V. The onset front propagates upward with a non constant velocity. The initial velocity of onset front decreases with voltage.} 
\label{period}
\end{center} \end{figure}

The intermittence does not always appear directly after the onset of the flow, but it always stops before the end of the discharge (see figure \ref{all}.c). One period of the intermittence can be separated into three phases (see figure \ref{period}). At the beginning, the heap is at rest everywhere, except near the outlet, where the flow is at its lowest value. A onset front propagates upward and reaches the top of the heap after a time $T_1$. This front doesn't propagate at a constant velocity : it starts at a velocity $v_1$ and reaches the top at a velocity $v_2$. The initial velocity $v_1$ decreases with voltage. It is of the approximately 2.6 m/s at 300V and 0.5 m/s at 1500V. The velocity $v_2$ is of the order of 5 m/s. The entire heap is in motion during the time interval $T_2$, and afterwards the flow stops sharply. 

\begin{figure}[!ht]
\begin{center} 
\includegraphics[scale=0.6]{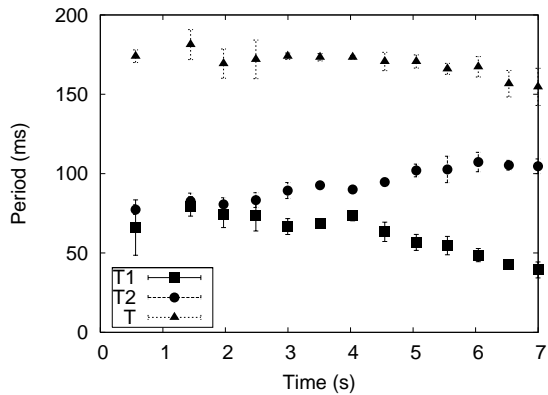} 
\caption{Typical dependence of the parameters $T_1$, $T_2$, $T$ with time during the silo discharge (V = 900 V). Each value is an average of three successive periods, the error bars width represents two standard deviation. After the last point of the curves, the flow continues but the intermittence disappears (see figure \ref{all}.(c)).} 
\label{T12T_t}
\end{center} \end{figure}

A typical evolution of the parameters $T_1$, $T_2$ and $T$ during the discharge is shown on figure \ref{T12T_t}. The period of onset $T_1$ slowly decreases with time $t$. It is not surprising : when the level of the heap decreases, the onset front needs less time to reach the top. The time $T_2$ increases with time. The ratio $T_2/T$ also increases with time. Knowing that the outlet mass flow is constant with time, this means that the mass flow at the top of the heap must decrease with time to allow mass conservation. In order to test if this evolution was due to the decrease of the packing height with discharge or to some aging, we reproduced the experiment with a filled hopper at the top of the silo. By this way, we maintained the level of the heap at its initial value. The different characteristic times evolve more slowly in this situation, showing that the evolution with time of the parameters is mainly due to the decrease of the height of the heap.

\begin{figure}[!ht]
\begin{center} 
\includegraphics[scale=0.6]{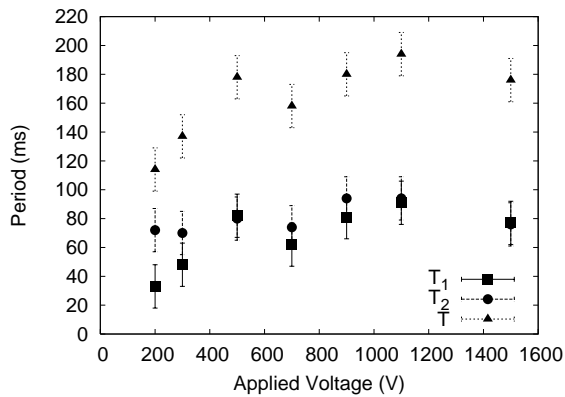} 
\caption{Dependence of the periods $T_1$, $T_2$, $T$ with voltage at the start of the flow.}
\label{T12T_V}
\end{center} \end{figure}

Figure \ref{T12T_V} exhibits the evolution of the characteristic times $T_1$, $T_2$ and $T$ with voltage. The time of onset $T_1$ increases with voltage. This can reflect the decrease of the initial velocity $v_1$. The times $T_2$ and $T$ also increase with voltage. However, the ratio $T_2$/$T$ decreases with voltage. Knowing that the mass flow is constant with voltage, this means that the instantaneous mass flow at the top of the heap must increase with voltage to fulfill mass conservation.

\section{Discussion}\label{effects}

In this section, we will discuss the flow behavior and the intermittences. To explain our observations, we must consider how an electric field acts on a dielectric granular material. Granular materials exhibit special conductive properties. The conduction of metallic granular materials has been studied by Dorbolo and coworkers. Ohm's law is not applicable to these kinds of materials and hysteretic behavior has been observed \cite{dorbolo}. Creyssels and coworkers suggested that electrical currents are concentrated in a collection of paths that are linked to the stress network \cite{creyssels}. Glass beads are sometimes called semi-insulating materials \cite{colver}. They are an insulator with a bulk  resistivity of at least 10$^{10}$$\Omega$m, but when the relative humidity of the ambient atmosphere is sufficiently high, they have a significant surface conductivity. This is responsible for many electrical properties of dielectric granular materials. High electric fields have already been used to modify the rheological properties of dielectric granular materials. Chen and coworkers applied a non-homogeneous electric field at the outlet of vertical pipes \cite{chen}. This affected the outlet mass flow and induced intermittences. The effects are stronger when the grains are in contact with the electrodes. This was applied to tune the flow of powders \cite{law}, to stabilize fluidized beds for heterogeneous catalytic reactions and solid gas reactions \cite{melcher}, or to avoid the segregation \cite{eow}. Robinson and Jones showed that the yield shear stress and maximal angle of stability of glass beads were greatly affected by the presence of electric fields \cite{Robinson_yield,Robinson}. They defined an electric stress which increases with voltage. Metayer studied the effect of electric fields on static and dynamic angles of repose of glass beads confined between two vertical electrode plates \cite{metayer}.

\subsection{Electrically induced interactions}\label{effect}

In this subsection, we will show that electric fields can be used to the tune cohesion of dielectric granular materials. This cohesion is anisotropic, and tends to form linear aggregates along the direction of the applied electric  field. Four effects could take place :

\begin{enumerate}

\item Dielectric grains can polarize and interact together with dipole/dipole interaction $F_{i,j}^{1}$.

\item Charged grains immersed in the applied electric field can experience a force $F_{i}^{2}$.

\item Charged grains can interact together with a force $F_{i,j}^{3}$.

\item The electrical current is responsible for an attractive interaction between grains, named electroclamping force $F_{i,j}^{4}$.

\end{enumerate}

In order to evaluate the strength of these effects, we will compare the mean force they induce on one single grain to the weight of this grain.

Let us first evaluate the interaction between induced dipoles. We can first consider one single dielectric sphere of diameter $d$ and relative permittivity $\epsilon_{r}$ immersed in a homogeneous electric field $\overrightarrow{E_0}$. The electric field resulting from the polarization of the sphere can be calculated analytically \cite{jackson}. Outside the sphere, it has the same form as the electric field produced by a dipole moment of value:
\begin{equation}
\vec{p} = \frac{\pi}{2}  \left ( \frac{ \epsilon_{r}-1}{\epsilon_{r}+2} \right ) \epsilon_{0} d^{3} \overrightarrow{E_{0}},
\label{dipole_moment}
\end{equation} where $\epsilon_{0}$ denotes the vacuum permittivity. We can use this expression to evaluate the interaction between two polarized dielectric spheres in contact. Let us call $\theta$ the angle between the orientation of the spheres and the electric field. If we consider that the grains have the same interaction as two dipoles given by equation (\ref{dipole_moment}), the force $F_{i,j}^{1}$ experienced by the grain $i$ in contact with the grain $j$ will be given by :

\begin{equation}
\overrightarrow{F_{i,j}^{1}} = \frac{1}{4 \pi \epsilon_0} \frac{3 {|\vec{p}|}^{2}}{d^4} ((1-3\cos^2 \theta)\overrightarrow{e_{i,j}} - 2\cos\theta \sin\theta \ \overrightarrow{e'_{i,j}})
\end{equation}

\label{F1} where $\overrightarrow{e_{i,j}}$ is the unit vector pointing from the center of $j$ to the center of $i$ and $\overrightarrow{e'_{i,j}}$ is the basis vector associated with the $\theta$ coordinate. This interaction is characterized by a Bond number $B_0^{1}=F_{i,j}^{1}/mg$, where $mg$ is the weight of one grain. We can evaluate this number using $\rho$ = 2500 kg/m$^{3}$ (density of silica glass), $d$ = 74 $\mu$m, $\epsilon_{r}=3.8$ (amorphous silica glass), $E_0$ =1900 V/ 6 mm = 316 kV/m). If the grains are aligned in the direction of the electric field ($\theta$ = 0), the interaction is attractive and

\begin{equation}
B_0^{1} \approx 2
\end{equation}

The other effects that we consider result from the superficial conductivity of the glass beads. In this work, we expect to have negligible cohesion between charged grains due to the triboelectric effect. At a relative humidity of 43 $\%$RH, the charge rapidly relaxes by surface conductivity. However, as grains are in contact with electrodes, they can acquire a charge. A grain $i$ carrying a charge $q_i$ and immersed in an electric field $\overrightarrow{E_0}$ will experience a force:

\begin{equation}
\overrightarrow{F_{i}^{2}} = q_i\overrightarrow{E_0}
\end{equation} We performed our experiment in conditions similar to the ones of Howell and co-workers \cite{aranson}. Using similar grains, applied electric field $E_0$ and hygrometry, they could charge grains sufficiently to have force $q_i E_0 > mg$ in a time short compared to the duration of our experiment. The question is: Is the quantity of charged grains sufficient for this effect to be efficient? In other words, is the mean ratio $B_0^2 = |q|_{mean} E_0/mg$ significant in our situation? In order to answer this question, we placed two vertical electrode plates at the outlet of the silo and we observed the trajectories of the grains falling between them. The horizontal acceleration of one grain is $qE/m$ while the vertical acceleration is $g$. By applying the same electric field $E_{0}$ inside the silo and between the electrodes, we can directly observe the ratio $F_i^{2}/mg$ by looking at the trajectories of the falling grains. This experiment showed that the application of the electric field induces charge of both signs on the grains. The ratio $|qE_{0}/mg|$ can be larger than one. Nevertheless, even for the largest values of the applied electric field, a large majority of the grains fall almost vertically and have a Bond number

\begin{equation}
B_0^2 \ll 1
\end{equation} This implies that the interaction of charged grains with the applied electric field does not play a significant role in our experiment.

One may also ask if the interaction between charged grains is significant. We consider the case of two touching dielectric spheres $i$ and $j$ of diameter $d$ and permittivity $\epsilon_{r}$ carrying charges $q_{i}$ and $q_{j}$ homogeneously distributed on their surface. Feng \cite{feng} used the finite element method to calculate a force :

\begin{equation}
\overrightarrow{F_{i,j}^{3}}=\frac{ \alpha(\epsilon_{r}) \left ( q_{i}^2+q_{j}^2 \right )-\beta(\epsilon_{r})  q_{i}q_{j} }{4 \pi \epsilon_{0} d^2} \overrightarrow{r_{i,j}},
\end{equation}
\label{bond3} where $\alpha(3.8) \approx 0.25$, $\beta(3.8) \approx 1.3$ and where $\overrightarrow{r_{i,j}}$ is the vector pointing from the center of $i$ to the center of $j$. If one considers that  $q_i E_0=q_j E_0=mg$, one finds a repulsive force of strength $F_{i,j}^{3}= 0.068 mg$. For $q_i E_0=mg$, $q_j=0$, one finds $F_{i,j}^{3} = 0.021 mg$. For $q_i=-q_j=mg/E_0$, one has an attractive force of strength 0.15 mg. However, we have seen that $|q|_{mean} \ll mg/E_0$. This interaction is thus characterized by a Bond number :

\begin{equation}
B_0^3  \ll 1,
\end{equation} showing that the interaction between grains charged by conductivity is not significant for our concern.

The last effect that has to be taken into account is the so-called electroclamping force. When a granular material constituted of glass beads is confined between two electrodes, a small electric current can pass through the packing due to adsorbed moisture \cite{drescher}. The density of current increases at the contact between grains, and thus the local electric field at the surface of the grain can be several orders of magnitude higher than the applied applied electric field \cite{moslehi}. The interaction between two grains can be calculated by integrating the electric stress over the surface of the grains. Different models attempted to describe this phenomenon quantitatively, despite the fact that several unknowns exist concerning the surface conductivity and the nature of the contact between grains \cite{ghadiri}. We will use the semi-empirical model proposed by Dietz and Melcher \cite{dietz}, which postulates that the conductivity is limited to the surface of the beads, that the grains deform according to the Hertz law, and that electrical breakdown of air occurs at the contact between grains. This model gives the following expression for the force between two grains :

\begin{equation}
 F_{i,j}^{4}= 0.415  \pi \epsilon_0 d^2 {E_{max}}^{0.8} (E_{0} cos \theta)^{1.2},
\end{equation} where $E_{max}$ = 3 10$^6$ V/m is the electric field for dielectric breakdown in air. Taking as before $E_{0}$ = 316 kV/m, $d$=74 $\mu$m and $\theta$=0, one finds a Bond number:
\begin{equation}
B_0^4 \approx 7
\end{equation} The electroclamping effect could then be relevant for our purpose.

The preceding discussion is summarized in Table \ref{tableau}. We considered four effects, two of them are significant for our purpose: the interaction between induced dipoles and the electroclamping force. Both tend to create linear clusters along the direction of the electric field. We can conclude that the application of a high electric field on a dielectric granular material can be responsible for a tunable anisotropic cohesion.

\begin{table}

\begin{tabular}{|c|c|c|}
\hline
Interaction & Notation & Bond $B_0$\\
\hline
Induced dipoles & $F_{i,j}^{1}$ & 2 \\ 
Grain/Field & $F_i^{2}$ & $\ll 1$ \\ 
Grain/Grain & $F_{i,j}^{3}$ & $\ll1$ \\ 
Electroclamping & $F_{i,j}^{4}$ & 7 \\ 
\hline
\end{tabular}
\caption{Four effects could play a role when applying an electric field on a granular material. Only the induced dipoles interaction and the electroclamping effect are efficient in our case.}
\label{tableau}
\end{table}

\subsection{Effect on the flow}\label{mass_effect}

The funnel to rathole transition described in section \ref{geometry} can be compared with experimental results of Robinson and Jones  \cite{Robinson}. These authors applied an electric field parallel to the axis of rotation of a rotating drum. They measured the maximum angle of stability of a packing of glass beads before the initiation of the flow. This angle increases with voltage up to 90\degre. They interpret this as an increase of the yield shear stress of the material. This can be compared to the slope of the walls that we observe in the rathole situation. The stabilization of an arch and the blockage of the flow for the highest voltage values could be viewed as the continuity of the funnel to rathole transition.

Our experiments have shown that the electric field doesn't affect the outlet mass flow. This is surprising for two reasons.

First, we report a transition from funnel flow to rathole flow, as well as the blockage of the flow. It is surprising to see that the static part of the heap is greatly affected by electric field, but that the flowing part isn't. However, it is not the first time that this contrast is reported. Metayer observed glass beads flowing between two parallel plate electrodes \cite{metayer}. He applied some voltage, imposed some inlet flow $Q$ and measured the dynamic angle of repose. For the smallest values of $Q$, this angle increased with voltage. But the effect of the electric field decreased for higher $Q$ values. Such dissimilarities between static and flowing heaps were also reported when applying a magnetic field on steel beads \cite{lemaire}.

The mass flow behavior is also surprising because the flow through the silo is expected to decrease with cohesion. This is in deep contrast with the results of Lumay and Vandewalle \cite{geoffroy2}. They applied both vertical and horizontal homogeneous magnetic field on ferromagnetic powder flowing in a hopper. In both cases, they observed a decrease of flow as a power 2 of the applied field \cite{geoffroy2}.

We can however use a schematic picture to justify the difference between our results. As depicted in section \ref{effect}, the electric field tends to create linear clusters aligned in the direction $z$ of the outlet slit. Let us consider a granular material with clusters aligned in the direction $z$. If one applies a shear stress, the cohesion can create a resistance to the gradient of velocity. For a shear stress $\sigma_{x,z}$ or $\sigma_{y,z}$, some cohesive links need to be broken for the material to shear. On the contrary, for a shear stress $\sigma_{x,y}$, no cohesive links will need to be broken for the material to shear. The clusters will be able to flow without any resistance. In our case, the clusters are aligned in the direction of the outlet slit and do not need to be broken to allow the flow through the silo.

\subsection{Effect on the intermittences}\label{mechanism}

In section \ref{spatio}, some spatiotemporal analysis was used to describe the intermittences. In this section, we suggest an interpretation of the phenomenon.

\begin{figure}[!ht]
\begin{center} 
\includegraphics[scale=0.5]{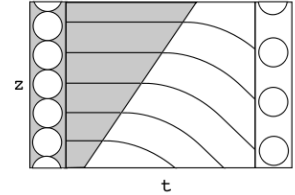} 
\caption{The upward propagation of a onset front induces a decrease of the density. In this picture, the density is linked to the vertical distance between trajectories.}
\label{density}
\end{center} \end{figure}

Let us consider one period of intermittency. At the beginning, the heap is at rest. The grains near the outlet start to flow, and the border between the static and flowing part of the heap propagates upwards. We call this discontinuity of acceleration the onset front. It must induce a dilatation of the moving part of the heap (see figure  \ref{density}).

The intermittence could arise because the electric field acts differently on static and dynamic part of the heap. As described in the previous section, the yield shear stress of a static packing increases with voltage \cite{Robinson_yield,Robinson}, but the effect of the electric field decreases when the flow increases \cite{metayer}. In our system, the electric field must have a greater effect during the onset front propagation. When some voltage is applied, the onset front propagates more slowly and a strong dilatation effect might occur. After the onset, the material is poorly affected by the electric field. The velocity field might not be be stationary. A compression could occur, and if the material reaches some critical density, the material must block. The cycle is repeated.

\section{Conclusion}\label{conclu}

We have studied the effect of an electric field on a dielectric granular material flowing through a silo. We have seen that electric field can induce some cohesion. Both the electroclamping force and the interaction between induced dipoles can be responsible for an anisotropic interaction between grains. This interaction tends to form linear clusters aligned in the direction of the applied electric field. Mass flow, flow geometry, and flow dynamics have been observed.

The mass flow was not affected by the electric field but the blockage occurred above a critical voltage. A transition from funnel flow to rathole was observed. This transition was of probabilistic nature : different kinds of flow geometry can occur at a given voltage. We observed an intermediate state between the funnel flow and the rathole flow. We called it the "avalanche regime". We observed a complete blockage of the heap after some critical voltage. The application of an electric field increased the probability to have an intermittent flow. Transition from continuous flow to intermittent flow was of a probabilistic nature and it was independent of the funnel to rathole transition.

The anisotropy of the cohesion might explain the independence of mass flow with voltage.  As already observed in similar experiments, the static part of the heap is more affected by the field than the flowing part. This might be the cause of the intermittence.

\section*{Acknowledgements} This work has been supported by INANOMAT project (Grant No. IAP P6/17) of the Belgian Science Policy. We would like to thank S. Dorbolo, F. Ludewig, T. Gilet, P. G\'erard and J.C. Remy for their help and valuable discussions.

\end{document}